\documentstyle[graphicx,aps,amstex,epsf]{revtex} \begin{document}
\textheight 23 cm 
\topmargin -1 cm 
\textwidth 15 cm 
\setlength{\baselineskip}{0.333333in}
\draft \preprint{ } 
\title{ Nonlinear coupling between scissors modes of a
Bose-Einstein condensate} 
\author{U.  Al Khawaja and H.  T.  C.  Stoof}
\address{Institute for Theoretical Physics, Utrecht University, Leuvenlaan
4, 3584 CE Utrecht, The Netherlands} 
\date{\today} 
\maketitle

\begin{abstract} We explore the nonlinear coupling of the three scissors
modes of an anisotropic Bose-Einstein condensate.  We show that only when
the frequency of one of the scissors modes is twice the frequency of
another scissors mode, these two modes can be resonantly coupled and a down
conversion can occur.  We perform the calculation variationally using a
gaussian trial wave function.  This enables us to obtain simple analytical
results that describe the oscillation and resonance behaviour of the two
coupled modes.  
\end{abstract}


\section{Introduction} Similar to monopole and quadrupole breathing modes
of a gaseous Bose-Einstein condensate
\cite{{jin},{kett},{stringari_old},{singh},{burnett},{castin},{perez}},
scissors modes were first studied theoretically \cite{{stringari},{pires}}
and subsequently observed experimentally \cite{{foot_first}}.  The scissors
modes are, however, rather special since they directly manifest the
superfluid behavior of these atomic gases.  Moreover, the recent
experimental studies appear to show a resonance behavior between two
coupled scissors modes \cite{foot_beliaev}.  From a theoretical point of
view this is interesting because a linear-response calculation can neither
account for the coupling nor for the resonance behavior
\cite{{stringari},{pires}}.  Therefore, a first step towards an explanation
of these experimental observations is to perform a calculation that goes
beyond linear-response theory and accurately takes into account the
mean-field interaction that couples the scissors and breathing modes.  In
this paper we present a simple variational method for calculating the
frequencies of these various modes and their couplings beyond the
linear-response.  We perform our calculation at zero temperature and
therefore do not consider the damping rates of the scissors modes
\cite{{foot_temp},{eugen}}.

The main idea behind our method is to use a time-dependent gaussian {\it
ansatz} for the groundstate wave function to derive the equations of motion
of the breathing modes and the scissors modes.  Then we expand the
resulting equations of motion in deviations from equilibrium.  In first
order, {\rm i.e.}, linear-response, we recover the expected uncoupled set
of equations \cite{{stringari},{pires}}.  The second-order calculation
produces a set of coupled equations which show that we need to excite all
three scissors modes in order to get a nonzero coupling.  At higher orders
we, however, find that we only need to excite two modes to get coupling.
Furthermore, we actually find under certain conditions a resonance behavior
between these two modes.

The layout of the paper is as follows.  First, we rederive in Sec.
\ref{freqs-section} the frequencies of the scissors modes in the
linear-response limit.  In Sec.  \ref{heigher} we extend the calculation
first to second, and then also to higher orders, which ultimately lead to a
resonant coupling.  In Sec.  \ref{solution} we solve the equations of
motion analytically near the resonance using an envelope function approach.
In Sec.  \ref{conclusion} we end with a discussion of our results.

\section{Frequencies of the scissors modes} \label{freqs-section} We start
by considering a Bose-Einstein condensate trapped by the following harmonic
potential \begin{equation} V({\bf r})=
{1\over2}m(\omega_{x}^{2}x^{2}+\omega_{y}^{2}y^{2}+\omega_{z}^{2} z^{2})
\label{v}, \end{equation} where $\omega_{x}$, $\omega_{y}$, and
$\omega_{z}$ are the angular frequencies of the trap, and $m$ is the atomic
mass.  A scissors mode in a Bose-Einstein condensate is associated with an
irrotational flow with a velocity field of the form ${\bf v}({\bf
r})\propto \mbox{\boldmath $\nabla$}(xy)$, if the motion is taking place in
the $xy$-plane \cite{stringari}.  Similar expressions hold for the two
other Cartisian planes.  These kind of modes can be excited by a sudden
rotation of the equilibrium axes of the trap.  To such a perturbation the
condensate will respond by oscillating around the new equilibrium axes.
For example, to excite a scissors mode in the $xy$-plane, we rotate the $x$
and $y$-axes of the trap slightly around the $z$-axis.  If the angle of
rotation is sufficiently small, the scissors mode can be approximated by a
simple oscillation of the condensate around the new equilibrium axes.  On
the other hand, if the axes change through a large angle this method
excites the $m=2$ quadrupole mode, where $m$ labels the projection of the
angular momentum along the axis of symmetry.  The maximum angle for which
the scissors mode is defined increases with deformation of the trap
\cite{foot_private}.

To account for all three scissors modes in the three Cartisian planes we
employ the following trial function for the condensate order parameter
\begin{equation} \psi({\bf r},t)=
A(t)\exp\left(-b_{x}(t)x^{2}-b_{y}(t)y^{2}-b_{z}(t)z^{2}-c_{xy}(t)xy-c_
{xz}(t)xz-c_{yz}(t)yz\right) \label{trial}, \end{equation} where $b_{i}$
and $c_{ij}$, are complex time-dependent variational parameters and
\begin{equation} A={\sqrt{N}2^{1/4}\over\pi^{3/4}}
\sqrt[4]{c_{xy,r}c_{xz,r}c_{yz,r}+4b_{x,r}b_{y,r}b_{z,r}-(b_{z,r}c_{xy,r}^2+b
_{y,r}c_{ xz,r}^2+b_{x,r}c_{yz,r}^2)} \label{a}.  \end{equation} This value
of the prefactor $A(t)$ guarantees the normalization of the square of the
wave function $\psi({\bf r},t)$ to the total number of condensed atoms $N$.
Here $b_{i,r}$ and $c_{ij,r}$ are the real parts of $b_{i}$ and $c_{ij}$,
respectively.  The first set of parameters, $b_{i}$, give rise to the
well-studied breathing modes which, for axially symmetric traps, are called
the monopole and quadrupole modes depending on the value of $m$ being equal
to zero or two, respectively.  The parameters $c_{ij}$ on the other hand
determine the three scissors modes.  The equations of motion for these
variational parameters can be derived from the lagrangian \begin{equation}
L[\psi,\psi^*]={1\over2}i\hbar\int{ {\rm d}{\bf r} \left(\psi^*({\bf
r},t){\partial\psi({\bf r},t)\over\partial t} -\psi({\bf
r},t){\partial\psi^*({\bf r},t)\over\partial t}\right)} -E[\psi,\psi^*]
\label{lagrangian}, \end{equation} where $E[\psi,\psi^*]$ is the usual
Gross-Pitaevskii energy functional given by \begin{equation}
E[\psi,\psi^*]=\int d{\bf r} \left[ {\hbar^2\over 2m}|{\mbox{\boldmath
$\nabla$}}\psi({\bf r},t)|^2 +V({\bf r})|\psi({\bf r},t)|^2
+{1\over2}T^{2B}|\psi({\bf r},t)|^4 -\mu|\psi({\bf r},t)|^2 \right]
\label{functional}.  \end{equation} Here $T^{2B}$ is the two-body
$T$-matrix, which for the atomic Bose-Einstein condensates of interest is
related to the $s$-wave scattering length $a$ through $T^{2B}=4\pi
a\hbar^2/m$.

Inserting our trial wave function into the lagrangian and scaling
frequencies with ${\bar \omega}=(\omega_{x}\omega_{y}\omega_{z})^{1/3}$ and
lengths with ${\bar a}=\sqrt{\hbar/m{\bar\omega}}$, it takes the
dimensionless form \begin{eqnarray} L[b,c]/N&=&\left(\alpha_{x}{\dot
b}_{x,i}+\alpha_{y}{\dot b}_{y,i}+\alpha_{z}{\dot
b}_{z,i}\right)/Q\\\nonumber &-&{1\over2}\left[
\alpha_{x}(4|b_{x}|^2+|c_{xy}|^2+|c_{xz}|^2)\right.\\\nonumber
&+&\alpha_{y}(4|b_{y}|^2+|c_{xy}|^2+|c_{yz}|^2)\\\nonumber
&+&\left.\alpha_{z}(4|b_{z}|^2+|c_{xz}|^2+|c_{yz}|^2)\right]/Q\\\nonumber
&-&{1\over2}\left[\alpha_{x}\omega_{x}^2+\alpha_{y}\omega_{y}^2+\alpha_{z
}\omega_{z}^2\right]/Q\\\nonumber &-&{1\over2\sqrt{\pi}}\gamma\sqrt{Q}
\label{scaled lagrangian}, \end{eqnarray} where $Q=2\pi^{3}A^4/N^2$,
$\alpha_{x}=4b_{y,r}b_{z,r}-c_{yz,r}^2$,
$\alpha_{y}=4b_{x,r}b_{z,r}-c_{xz,r}^2$,
$\alpha_{z}=4b_{x,r}b_{y,r}-c_{xy,r}^2$, and the dot corresponds to a time
derivative.  In addition $\gamma=Na/{\bar a}$ is the dimensionless
parameter that represents the strength of the mean-field interaction.
Minimizing the lagrangian with respect to the 12 variational parameters, we
get a set of 12 coupled equations of motion.  The resulting equations of
motion are rather lengthy and complicated.  A significant simplification
takes place if we expand these equations in the deviation of the
variational parameters from their equilibrium values, {\rm i.e.}, in
$\delta b_{i}(t)=\exp{(-i\omega t)}(b_{i}-b_{i}^{(0)})= \exp(-i\omega
t)\delta b_i$ and $\delta c_{ij}(t)=\exp{(-i\omega
t)}(c_{ij}-c_{ij}^{(0)})= \exp(-i\omega t)\delta c_{ij}$, where $\omega$ is
the yet unknown eigenfrequency of the modes, and $b_{i}^{(0)}$ and
$c_{ij}^{(0)}$ denote the equilibrium values of $b_{i}$ and $c_{ij}$,
respectively.  These latter values can simply be obtained by setting the
time derivatives in the equations of motion to zero.  For large condensates
in the so-called Thomas-Fermi regime we can ignore contributions from the
kinetic energy \cite{pethick} and the equilibrium variational parameters
take the simple form \begin{eqnarray}
b_{i}^{(0)}&=&({\sqrt{\pi}\over8\gamma})^{2/5} \omega_{i}^{2},\nonumber\\
c_{ij}^{(0)}&=&0 \label{equilibrium bees}.  \end{eqnarray} It is required
that $\gamma\gg 1$ for the last equation to be valid.  To first order in
the deviations the equations of motion read simply \begin{eqnarray} {\bf
M}\cdot{\bf P}=0 \label{2ndorder}, \end{eqnarray} where the vector
\begin{math}{\bf P}=(\delta b_{x,r},\delta b_{y,r},\delta b_{z,r}, \delta
b_{x,i},\delta b_{y,i},\delta b_{z,i}, \delta c_{xy,r},\delta
c_{xz,r},\delta c_{yz,r}, \delta c_{xy,i},\delta c_{xz,i},\delta c_{yz,i})
\end{math} contains all the possible fluctuations, and the matrix ${{\bf
M}}$ is given by \begin{equation} {{\bf M}}=\left( \begin{array}{cc} {{\bf
M}}^{\rm breathing}&0\\ 0&{{\bf M}}^{\rm scissors} \end{array} \right)
\label{x1}, \end{equation} where ${{\bf M}}^{\rm breathing}$ and ${\bf
M}^{\rm scissors}$ are given explicitly in Appendix \ref{app}.  It is clear
from the last equation that to linear order the breathing modes and
scissors modes are uncoupled.  The dispersion relation of these modes can
be obtained by setting the determinant of ${{\bf M}}$ to zero.  This
results in \begin{eqnarray}
&(&\omega^2-\Omega_{xy}^2)(\omega^2-\Omega_{xz}^2)(\omega^2-\Omega_{yz}^2
)\nonumber\\ \times
&(&\omega^6-3\omega_a^2\omega^4+8\omega_b^4\omega^2-20\omega_c^6) =0
\label{dispersionlinear}, \end{eqnarray} where
$\Omega_{xy}=\sqrt{\omega_x^2+\omega_y^2}$,
$\Omega_{xz}=\sqrt{\omega_x^2+\omega_z^2}$,
$\Omega_{yz}=\sqrt{\omega_y^2+\omega_z^2}$,
$\omega_a^2=\omega_x^2+\omega_y^2+\omega_z^2$,
$\omega_b^4=\omega_x^2\omega_y^2+\omega_x^2\omega_z^2+\omega_y^2\omega_z^2$,
and $\omega_c^6=(\omega_x\omega_y\omega_z)^2$.  The zeros of the first
three factors in the left-hand side of the last equation give the
frequencies of the scissors modes, whereas the zeros of the sixth-order
polynomial give the frequencies of the breathing modes.

So far we have not produced a new result, since these frequencies have been
calculated previously \cite{{stringari},{pires},{stringari_modern}}.
However, we have demonstrated a new way of calculating the frequencies of
scissors modes in a relatively simple manner.  The most important part of
this paper is therefore contained in the next section, where we consider
also the nonlinear effects produced by the mean-field interaction.

\section{Beyond linear response} \label{heigher} In this section we
consider the equations of motion for the variational parameters by taking
into account several higher-order terms in the deviations $\delta b_i$ and
$\delta c_{ij}$.  We have calculated these equations analytically up to
second order, but they turn out to be rather lengthy and contain terms that
couple the breathing modes and the scissors modes.  As we discuss below, in
the present experiments with axially symmetric traps the coupling of the
scissors modes with the quadrupole mode is always of importance, but we
leave the treatment of this more complicated situation to future work.  For
simplicity, therefore, we focus here on triaxial traps, in which case we
can ignore the breathing modes.  This is achieved by simply setting $\delta
b_{i}=0$ in the full equations of motion.

Up to second order, the remaining 6 first-order equations for $\delta
c_{xy,r}\dots\delta c_{yz,i}$ can be reduced to 3 second-order equations
for $\delta c_{xy,r}$, $\delta c_{xz,r}$, and $\delta c_{yz,r}$ by
eliminating the imaginary parts.  We find in detail \begin{equation}
{\delta \ddot c}_{xy,r}+\Omega_{xy}^2\delta c_{xy,r}
+({64\over\pi})^{1/5}\gamma^{2/5}
\left({\omega_x\omega_y\over\omega_z^{4}}\right)^{2/5} \left[
-\Omega_{xy}^2\delta c_{xz,r}\delta c_{yz,r}
+2\left({\omega_x\omega_y\over\Omega_{xz}\Omega_{yz}}\right)^2 {\delta
\dot{c}_{xz,r}}{\delta \dot{c}_{yz,r}} \right] \label{c1}, \end{equation}
\begin{equation} {\delta \ddot{c}}_{xz,r}+\Omega_{xz}^2\delta c_{xz,r}
+({64\over\pi})^{1/5}\gamma^{2/5}\left({\omega_x\omega_z\over\omega_y^{4}}\right)^{2/5}
\left[ -\Omega_{xz}^2\delta c_{xy,r}\delta c_{yz,r}
+2\left({\omega_x\omega_z\over\Omega_{xy}\Omega_{yz}}\right)^2 {\delta
\dot{c}_{xy,r}}{\delta \dot{c}_{yz,r}} \right] \label{c2}, \end{equation}
and \begin{equation} {\delta \ddot c}_{yz,r}+\Omega_{yz}^2\delta c_{yz,r}
+({64\over\pi})^{1/5}\gamma^{2/5}
\left({\omega_y\omega_z\over\omega_x^4}\right)^{2/5} \left[
-\Omega_{yz}^2\delta c_{xy,r}\delta c_{xz,r}
+2\left({\omega_y\omega_z\over\Omega_{xy}\Omega_{xz}}\right)^2 {\delta
\dot{c}_{xy,r}}{\delta \dot{c}_{xz,r}} \right] \label{c3}.  \end{equation}
These equations show that if we ignore the second-order terms we recover
our previous three uncoupled scissors modes with frequencies $\Omega_{xy}$,
$\Omega_{xz}$ and $\Omega_{yz}$.  It is interesting to observe that the
coupling terms couple the three scissors modes such that if one mode is not
excited the other two modes will not be coupled.  In other words, to get a
nonzero coupling up to this order, the three modes must be excited
simultaneously.  We believe that this important result is not an artefact
of the gaussian approximation, but also holds for an exact calculation
using the Bogoliubov theory.

For higher-order couplings the last conclusion is no longer true.  Two
modes can then be coupled, even when the third is not excited.  From now
on, therefore, we assume without loss of generality that only the $\delta
c_{xy}$ and $\delta c_{xz}$ modes are excited, while $\delta c_{yz}=0$.
Moreover, to investigate the possibility of a resonant coupling between
these two modes, we have considered coupling terms up to ninth order.
Similar to the second-order case explained above, the equations of motion
can thus be expressed as \begin{equation} {\delta \ddot
c}_{xy,r}+\Omega_{xy}^2\delta c_{xy,r} +\sum_{j+k+l+m\leq9}
\alpha_{jklm}(\delta c_{xy,r})^j({\delta \dot c}_{xy,r})^k (\delta
c_{xz,r})^l({\delta \dot c}_{xz,r})^m \label{c1_general}, \end{equation}
\begin{equation} {\delta \ddot c}_{xz,r}+\Omega_{xz}^2\delta c_{xz,r}
+\sum_{j+k+l+m\leq9} \beta_{jklm}(\delta c_{xz,r})^j({\delta \dot
c}_{xz,r})^k (\delta c_{xy,r})^l({\delta \dot c}_{xy,r})^m
\label{c2_general}, \end{equation} where $j,k,l,m=0,1,2,3$ and the sum
$j+k+l+m$ does not excced 9, which is the order up to which we have chosen
to expand the equations of motion.  The coefficients $\alpha_{jklm}$,
$\beta_{jklm}$ are given in terms of the trap parameters $\omega_{x}$,
$\omega_{y}$, $\omega_{z}$, and the interaction parameter $\gamma$.  A
resonance between these two modes takes place when the frequency of the
coupling terms is equal to the frequency of the zeroth-order term, {\rm
i.e.}, the first two terms.  The frequency of each coupling term in the
above summations is determined by substituting for ${\delta c_{ij}}$ and
${\delta \dot c_{ij}}$ their zeroth-order solutions.  Therefore, the latter
frequencies will be a certain linear combination of the zeroth-order
scissors modes frequencies.  Imposing the above resonance condition on each
coupling term thus results in a relation between the two scissors modes
frequencies $\Omega_{xy}$ and $\Omega_{xz}$.  Inspecting all coupling terms
up to the ninth order, we found a very small number of terms for which the
relation between $\Omega_{xy}$ and $\Omega_{xz}$ can be satisfied by real
values of $\omega_x$, $\omega_y$, and $\omega_z$.  Ultimately, we find only
a resonance when either \begin{equation} \Omega_{xy}=\Omega_{xz},
\end{equation} or \begin{equation} 2\Omega_{xy}=\Omega_{xz}, \label{cond2}
\end{equation} is satisfied.  The first resonance condition leads in the
experimentally relevant axially symmetric case, where
$\omega_x=\omega_y=\omega_z/\sqrt{\lambda}$, to a value for the anisotropy
ratio $\lambda$ that is equal to 1.  This clearly corresponds to a
spherically symmetric condensate.  Since in this case the scissors modes
are degenerate with the quadrupole breathing modes, which we have ignored
here, we focus from now on only on the second resonance condition.

It is interesting to mention that the second resonance condition is exactly
the same as the one observed experimentally \cite{foot_beliaev}.  For a
resonance of this kind the resonant coupling terms turn out to be of
seventh order, and the equations of motion in that case read
\begin{equation} {\delta \ddot c}_{xy,r}+\Omega_{xz}^2\delta c_{xy,r}
+\beta\; (\delta c_{xy,r})^3(\delta {\dot c}_{xy,r})^2(\delta c_{xz,r})^2
+\eta\; (\delta c_{xy,r})^5 (\delta\dot{c}_{xz,r})^2=0 \label{c1n},
\end{equation} \begin{equation} {\delta \ddot c}_{xz,r}+\Omega_{xz}^2\delta
c_{xz,r} +\alpha \;(\delta c_{xz,r})^3(\delta
c_{xy,r})^2(\delta\dot{c}_{xy,r})^2=0 \label{c2n}, \end{equation} if we
neglect all nonresonant terms.  Here $\alpha$, $\beta$ and $\eta$ are
functions of $\omega_x$, $\omega_y$, $\omega_z$, and $\gamma$ that are
given explicitly in Appendix \ref{app}.  It is important to note here that
our neglect of the nonresonant terms is justified when we are close enough
to resonance.  This is similar to the rotating-wave approximation known
from quantum optics.  We see that the coupling terms indeed lead to the
above-mentioned resonance condition, by inserting in them the zeroth-order
solutions, {\rm i.e.}, $\delta c_{xy,r}\propto\exp{(-i\Omega_{xy}t)}$ and
$\delta c_{xz,r}\propto\exp{(i\Omega_{xz}t)}$.  For example, the coupling
terms in Eq.  (\ref{c1n}) have a total frequency of
$3\Omega_{xz}-4\Omega_{xy}$.  Separating out $\delta
c_{xy,r}\propto\exp{(-i\Omega_{xz}t)}$ as a prefactor for the whole
equation, the coupling term oscillates thus as
$2\Delta=4\Omega_{xy}-2\Omega_{xz}$.  Therefore, a resonance takes place
when $\Delta=0$, {\rm{i.e.}}, when the condition in Eq.  (\ref{cond2}) is
met.  Similarly, the coupling term of Eq.  (\ref{c2n}) is also oscillating
with a frequency of $2\Delta$.

\section{Solution of the equations of motion near resonance}
\label{solution} Sufficiently close to resonance we can write $\delta
c_{xy}$ and $\delta c_{xz}$ as a product of two functions.  One of them
describes the slow envelope and the other the fast oscillation with the
uncoupled scissors-mode frequency.  In particular, we have \begin{equation}
\delta c_{xy}(t)=g(t)\exp{(i\Omega_{xy}t)} \label{geq}, \end{equation} and
\begin{equation} \delta c_{xz}(t)=f(t)\exp{(-i\Omega_{xz}t)} \label{feq},
\end{equation} where $g(t)$ and $f(t)$ are the slowly varying envelope
functions.  Substituting these expressions into Eqs.  (\ref{c1n}) and
(\ref{c2n}), ignoring second-order time derivatives of $f(t)$ and $g(t)$,
and then eliminating $g(t)$, we obtain the following equation for $f(t)$
\begin{equation} -(i{\ddot{f}}+2\Delta{\dot{f}})f+i\varepsilon
{{\dot{f}}}^2=0 \label{f2eq}, \end{equation} where \begin{equation}
\varepsilon={3\alpha\Omega_{xy}^3-4\beta\Omega_{xy}^2\Omega_{xz}-4\eta
\Omega_{xz}^3\over\alpha\Omega_{xy}^3}.  \end{equation} This equation has a
solution of the form \begin{equation} f(t)=\left[C_1+C_2\exp{(2i\Delta
t)}\right]^{1\over1-\varepsilon} \label{sol}, \end{equation} where $C_1$
and $C_2$ are two constants of integration that are determined by the
initial conditions.  Note that the relevant quantity here is $|f(t)|$,
which represents the actual envelope of the oscillation and is given by
\begin{equation} |f(t)|=\left[C_1^2+C_2^2+2C_1C_2\cos{(2\Delta
t)}\right]^{1\over2(1-\varepsilon)}.  \end{equation} In first instance we
might think that the real part of $f(t)$ is the relevant quantity.
However, in Eqs.  (\ref{geq}) and (\ref{feq}) we should in principle have
taken the real part of the right-hand side.  If we do that we automatically
are lead to the condition that $|f(t)|$ is the envelope of the oscillation.

For definiteness sake let us take the initial conditions
$f(0)=f_{r}(0)\equiv f_{0}$ and $\dot{f}(0)={\dot f}_{i}(0)\equiv{\dot
f}_{0}$, where $f_{r}(t)$ and $f_{i}(t)$ are the real and imaginary parts
of $f(t)$, respectively.  Physically, this set of initial conditions
corresponds to exciting the scissors modes in the $xz$ and the $xy$-planes
simultaneously.  This should be performed experimentally by initially
rotating the condensate in the $xz$-plane by an angle $\theta_0$ and around
the $z$-axis by an angle $\phi_0$ and then releasing the condensate.  The
initial angles $\theta_0$ and $\phi_0$ are related to the constants $f_{0}$
and ${\dot f}_{0}$ by \begin{equation}
f_{0}=2|b_{x}^{(0)}-b_z^{(0)}|\cos{\theta_0}\;\sin{\theta_0},
\end{equation} \begin{equation} {\dot
f}_{0}={\alpha\Omega_{xy}^2\over\Omega_{xz}}
(2|b_{x}^{(0)}-b_z^{(0)}|\cos{\theta_0}\;\sin{\theta_0})^3
|b_{x}^{(0)}-b_y^{(0)}|\cos{\phi_0}\;\sin{\phi_0}.  \end{equation} With
these initial conditions the constants $C_1$ and $C_2$ are given by
\begin{equation} C_1=f_{0}^{1-\varepsilon}+{{\dot f}_{0}
f_{0}^{-\varepsilon}(1-\varepsilon)\over 2\Delta }, \end{equation}
\begin{equation} C_2=-{{\dot f}_{0}f_{0}^{-\varepsilon}(1-\varepsilon)\over
2\Delta }.  \end{equation} Using these expressions and the experimental
parameters from Ref.  \cite{foot_first}, we give in Fig.  \ref{2a} the real
part of $\delta c_{xz}(t)$.  This clearly shows how the energy is being
exchanged between the two modes.

An interesting property of Eq.  (\ref{f2eq}) is that exactly on resonance,
{\rm i.e.}, $\Delta=0$, its solution becomes nonoscillatory.  Indeed we
find that in this case the solution is \begin{equation} f(t)=(C_3+C_4
t)^{1\over1-\varepsilon}, \label{sol_dec} \end{equation} where again $C_3$
and $C_4$ are constants that are determined by the initial conditions.  For
the above initial conditions this is a decreasing function in time.
Physically, this means that, unlike the case of an off-resonant
oscillation, it takes an infinite time for the energy that is transferred
from the scissors mode in the $xz$-plane to the scissors mode in the
$xy$-plane to get back to the mode in the $xz$-plane.  In Fig.  \ref{2b} we
show this resonance behavior.  Finally, we can show that in the limit
$\Delta\rightarrow0$ the oscillatory solution given in Eq.  (\ref{sol})
reduces to the nonoscillatory one at resonance given by Eq.
(\ref{sol_dec}).

\section{Summary and Conclusion} \label{conclusion} We have explored the
role of the mean-field interaction in coupling the three scissors modes of
a Bose-Einstein condensate.  A variational approach with a gaussian trial
wave function, that contains a number of variational parameters describing
the scissors modes, provides a relatively simple way in which we can
extract the main features of this coupling.  To first order in the
deviations in the variational parameters from their equilibrium values we
reproduce the correct frequencies of the scissors modes.  To second order
we show that it is not possible to have two modes that are coupled if the
third mode is not excited.  Instead, all three modes need to be excited for
nonlinear dynamics to occur.

At higher orders we find that it is possible to couple only two modes.  In
this case we find a resonance behavior if $2\Omega_{xy}=\Omega_{xz}$ or
$\Omega_{xy}=\Omega_{xz}$.  Up to the nineth order in the deviations, these
are the only two cases of resonance that we have found.  Close to resonance
the equations of motion have been solved exactly using an envelope
approach.  The resulting dynamics is similar to a beating between two modes
with a beating frequency $2\Delta=|2\Omega_{xy}-\Omega_{xz}|$.  We notice
that the observed resonance behavior \cite{foot_beliaev} occurs exactly for
the same condition that we have obtained, namely
$2\Omega_{xy}=\Omega_{xz}$.  Also the fact that we did not find any other
resonance condition up to the ninth order, indicates that the coupling
terms that lead to this resonance are also responsible for the experimental
resonance in the triaxial case.

Furthermore, it is important to note here that, quite generally, the
down-conversion process from one excitation quantum into two excitation
quanta with half the energy each, i.e., the so-called Beliaev damping
\cite{beliaev}, vanishes for the scissors modes.  This is so because of the
negative parity of the scissors modes.  In terms of fluctuations of the
anihilation operator $\hat{\psi} ({\bf r},t)$ given by $\hat{\varphi} ({\bf
r},t)=\hat{\psi}({\bf r},t)-\psi({\bf r},t)$, the Beliaev damping process
is accounted for by an interaction term proportional to \begin{math} \int
d{\bf r}\;{ \psi({\bf r},t)\hat{\varphi}^{\dagger} ({\bf
r},t)\hat{\varphi}^{\dagger} ({\bf r},t) \hat{\varphi}({\bf r},t)
}\end{math}.  Here $\psi({\bf r},t)\equiv<\hat{\psi}({\bf r},t)>$ is again
the condensate wave function.  Using our variational gaussian wave function
given by Eq.  (\ref{trial}), we clearly see that with only two scissors
modes present this integral vanishes, since the integrant is an odd
function.  We believe that this result is independent of our trial wave
function and also true within an exact approach \cite{foot_lower}.  In our
case the seventh-order coupling terms in Eqs.  (\ref{c1n}) and (\ref{c2n})
correspond to a quadratic collisional damping process, where three
excitation quanta decay into four excitation quanta with half the frequency
and one excitation quantum with the same frequency.  This is shown
schematically in Fig.  \ref{schem}.  This makes sense physically since this
is the lowest-order nonvanishing process if Beliaev damping is forbidden
and we are forced to apply the interaction term proportional to
\begin{math} \int d{\bf r}{ \hat{\varphi}^{\dagger} ({\bf
r},t)\hat{\varphi}^{\dagger} ({\bf r},t) \hat{\varphi}({\bf
r},t)\hat{\varphi}({\bf r},t) }\end{math} twice to accomplish the down
conversion.  Another point worth mentioning here is that in the experiments
the resonance is studied in most detail for an axially symmetric trap with
an anisotropy ratio close to $\lambda=\sqrt{7}$, when the $xy$-scissors
mode is degenerate with the quadrupole mode, which we have excluded in the
present calculation.  In that case we may argue that one scissors-mode
quantum can decay into either another scissors-mode quantum and a
quadrupole-mode quantum or into two quadrupole-mode quanta.  These are in
principle also Beliaev damping processes, which are however again forbidden
for similar reasons as before.  Nevertheless, a good understanding of these
experiments requires an analysis that includes not only the $xy$ and the
$xz$ scissors modes but also the quadrupole mode, because of the degeneracy
that occures in this case.  However, our theory should be directly
applicable to the experiments with a triaxial trap.  We hope that in the
future more detailed experiments of this kind will also be performed, to
make a direct comparison between our theory and experiment possible.

\section*{Acknowledgements} We would like to thank C.  J.  Foot and P.
Drummond for useful discussions.  This work is supported by the Stichting
voor Fundamenteel Onderzoek der Materie (FOM), which is financially
supported by the Nederlandse Organisatie voor Wetenschappelijk Onderzoek
(NWO).

 \newpage

\section*{Figure Captions} \begin{figure}[p] \begin{center}

\end{center} 
\caption{The
real part of $\delta c_{xz}$ showing two kinds of oscillation.  The one
with the larger frequency corresponds to the unperturbed scissors mode
oscillation with frequency $\Omega_{xz}$.  The slower oscillation is due to
the mean-field coupling between the scissors mode in the $xz$-plane and the
$xy$-plane.  The frequency of this oscillation is
$2\Delta=2|2\Omega_{xy}-\Omega{xz}|$.  The Bose-Einstein condensate
parameters used to make this plot are
$\omega_{x}=\omega_{y}=\omega_{z}/\sqrt{\lambda}=128$ Hz.  As in the case
of Ref.  [10], $\lambda=2.54$, and $N=10^4$ of $^{87}$Rb atoms.  The
initial conditions are $\theta_0=20\times\pi/180$ and $\phi_0=0.03\times
\pi/180$.  The latter was taken small to show that already such a small
perturbation in the $xz$-scissors mode is sufficient to initiate a
substantial coupling between this mode and the $xy$-mode.}  
\label{2a}
\end{figure}

\begin{figure}[p] \begin{center}

\end{center} \caption{The
envelope $|f(t)|$ that determines the energy transferred between the two
coupled scissors modes.  This set of curves shows that when approaching the
resonance condition $\Delta=|2\Omega_{xy}-\Omega_{xz}|=0$, the frequency of
the curve decreases until it becomes equal to zero on resonance.  The inset
shows this last behaviour for much larger times.  For axially symmetric
traps the resonance takes place at $\lambda=\sqrt{7}$.  The parameters used
for these plots are the same as those mentioned in Fig.  \ref{2a}.  }
\label{2b} \end{figure}

\begin{figure}[p] \begin{center}

\end{center} 
\caption{A
schematic figure representing three down-conversion processes.  a) Beliaev
damping where one excitation decays into two excitations with half the
frequency.  It is shown in the text that for two scissors modes this
process vanishes.  b) quadratic collisional damping in which three
excitations decay into one excitation of the same frequency and four
excitations with half the frequency.  According to the present work this is
the first nonvanishing down-conversion process.  c) Beliaev damping due to
down conversion from one scissors mode into another scissors mode and a
degenerate quadrupole mode, or into two quadrupole modes.  This is the
situation that corresponds to the most detailed experiments of Ref.  [11],
but the matrix elements of these proces again vanish.  } \label{schem}
\end{figure}

\newpage \section*{Figures}

\appendix \section{Coupling matrices and coefficients} \label{app} The
matrices ${{\bf M}}^{\rm breathing}$ and ${{\bf M}}^{\rm scissors}$ are
given by \begin{equation} {{\bf M}}^{\rm breathing}= \left(
\begin{array}{llllll} \frac{-3{\gamma }^{\frac{2}{5}}{{{\omega
}_x}}^{\frac{2}{5}}{{{\omega }_y}}^{\frac{2}{5}}{{{\omega
}_z}}^{\frac{2}{5}}} {2^{\frac{4}{5}}{\pi }^{\frac{1}{5}}}&- \frac{{\gamma
}^{\frac{2}{5}}{{{\omega }_x}}^{\frac{12}{5}}{{{\omega }_z}}^{\frac{2}{5}}}
{2^{\frac{4}{5}}{\pi }^{\frac{1}{5}}{{{\omega }_y}}^{\frac{8}{5}}} & -
\frac{{\gamma }^{\frac{2}{5}}{{{\omega }_x}}^{\frac{12}{5}}{{{\omega
}_y}}^{\frac{2}{5}}} {2^{\frac{4}{5}}{\pi }^{\frac{1}{5}}{{{\omega
}_z}}^{\frac{8}{5}}} &i \omega &0&0\\ - \frac{{\gamma
}^{\frac{2}{5}}{{{\omega }_x}}^{\frac{12}{5}}{{{\omega }_z}}^{\frac{2}{5}}}
{2^{\frac{4}{5}}{\pi }^{\frac{1}{5}}{{{\omega }_y}}^{\frac{8}{5}}} &
\frac{-3{\gamma }^{\frac{2}{5}}{{{\omega }_x}}^{\frac{22}{5}}{{{\omega
}_z}}^{\frac{2}{5}}} {2^{\frac{4}{5}}{\pi }^{\frac{1}{5}}{{{\omega
}_y}}^{\frac{18}{5}}}& - \frac{{\gamma }^{\frac{2}{5}}{{{\omega
}_x}}^{\frac{22}{5}}} {2^{\frac{4}{5}}{\pi }^{\frac{1}{5}}{{{\omega
}_y}}^{\frac{8}{5}}{{{\omega }_z}}^{\frac{8}{5}}} &0&\frac{i \omega
{{{\omega }_x}}^4}{{{{\omega }_y}}^4}&0\\ - \frac{{\gamma
}^{\frac{2}{5}}{{{\omega }_x}}^{\frac{12}{5}}{{{\omega }_y}}^{\frac{2}{5}}}
{2^{\frac{4}{5}}{\pi }^{\frac{1}{5}}{{{\omega }_z}}^{\frac{8}{5}}} & -
\frac{{\gamma }^{\frac{2}{5}}{{{\omega }_x}}^{\frac{22}{5}}}
{2^{\frac{4}{5}}{\pi }^{\frac{1}{5}}{{{\omega }_y}}^{\frac{8}{5}}{{{\omega
}_z}}^{\frac{8}{5}}}& \frac{-3{\gamma }^{\frac{2}{5}}{{{\omega
}_x}}^{\frac{22}{5}}{{{\omega }_y}}^{\frac{2}{5}}} {2^{\frac{4}{5}}{\pi
}^{\frac{1}{5}}{{{\omega }_z}}^{\frac{18}{5}}}&0&0&\frac{i \omega {{{\omega
}_x}}^4}{{{{\omega }_z}}^4} \\ -i \omega &0&0&- \frac{2^{\frac{4}{5}}{\pi
}^{\frac{1}{5}}{{{\omega }_x}}^{\frac{8}{5}}} {{\gamma
}^{\frac{2}{5}}{{{\omega }_y}}^{\frac{2}{5}}{{{\omega }_z}}^{\frac{2}{5}}}
&0&0 \\ 0&\frac{-i \omega {{{\omega }_x}}^4}{{{{\omega }_y}}^4}&0&0& -
\frac{2^{\frac{4}{5}}{\pi }^{\frac{1}{5}}{{{\omega }_x}}^{\frac{18}{5}}}
{{\gamma }^{\frac{2}{5}}{{{\omega }_y}}^{\frac{12}{5}}{{{\omega
}_z}}^{\frac{2}{5}}} &0 \\ 0&0&\frac{-i \omega {{{\omega }_x}}^4}{{{{\omega
}_z}}^4}&0&0& - \frac{2^{\frac{4}{5}}{\pi }^{\frac{1}{5}}{{{\omega
}_x}}^{\frac{18}{5}}} {{\gamma }^{\frac{2}{5}}{{{\omega
}_y}}^{\frac{2}{5}}{{{\omega }_z}}^{\frac{12}{5}}} \end{array} \right)
\label{m1b} \end{equation} and \begin{eqnarray} {{\bf M}}^{\rm scissors}=
\left( \begin{array}{llllll} - \frac{{\gamma }^{\frac{2}{5}}{{{\omega
}_x}}^{\frac{12}{5}}{{{\omega }_z}}^{\frac{2}{5}}} {2^{\frac{4}{5}}{\pi
}^{\frac{1}{5}}{{{\omega }_y}}^{\frac{8}{5}}} &0&0& \frac{\frac{i
}{2}\omega {{{\omega }_x}}^2}{{{{\omega }_y}}^2}&0&0\\ 0&- \frac{{\gamma
}^{\frac{2}{5}}{{{\omega }_x}}^{\frac{12}{5}}{{{\omega }_y}}^{\frac{2}{5}}}
{2^{\frac{4}{5}}{\pi }^{\frac{1}{5}}{{{\omega }_z}}^{\frac{8}{5}}} &0&0&
\frac{\frac{i }{2}\omega {{{\omega }_x}}^2}{{{{\omega }_z}}^2}&0\\ 0&0&-
\frac{{\gamma }^{\frac{2}{5}}{{{\omega }_x}}^{\frac{22}{5}}}
{2^{\frac{4}{5}}{\pi }^{\frac{1}{5}}{{{\omega }_y}}^{\frac{8}{5}}{{{\omega
}_z}}^{\frac{8}{5}}} &0&0& \frac{\frac{i }{2}\omega {{{\omega
}_x}}^4}{{{{\omega }_y}}^2{{{\omega }_z}}^2}\\ \frac{\frac{-i }{2}\omega
{{{\omega }_x}}^2}{{{{\omega }_y}}^2}&0&0& \frac{- { \frac{\pi }{2}
}^{\frac{1}{5}}{{{\omega }_x}}^{\frac{8}{5}} {{{\omega }_x}}^2 + {{{\omega
}_y}}^2 }{2{\gamma }^{\frac{2}{5}}{{{\omega }_y}}^{\frac{12}{5}} {{{\omega
}_z}}^{\frac{2}{5}}}&0&0\\ 0&\frac{\frac{-i }{2}\omega {{{\omega
}_x}}^2}{{{{\omega }_z}}^2}&0&0& \frac{- { \frac{\pi }{2}
}^{\frac{1}{5}}{{{\omega }_x}}^{\frac{8}{5}} {{{\omega }_x}}^2 + {{{\omega
}_z}}^2 }{2{\gamma }^{\frac{2}{5}}{{{\omega }_y}}^{\frac{2}{5}} {{{\omega
}_z}}^{\frac{12}{5}}}&0\\ 0&0&\frac{\frac{-i }{2}\omega {{{\omega
}_x}}^4}{{{{\omega }_y}}^2{{{\omega }_z}}^2}&0&0& \frac{- { \frac{\pi }{2}
}^{\frac{1}{5}}{{{\omega }_x}}^{\frac{18}{5}} {{{\omega }_y}}^2 + {{{\omega
}_z}}^2 }{2{\gamma }^{\frac{2}{5}}{{{\omega }_y}}^{\frac{12}{5}} {{{\omega
}_z}}^{\frac{12}{5}}} \end{array} \right).\\\nonumber \label{m1s}
\end{eqnarray} The coupling coefficients $\alpha$, $\beta$ and $\eta$,
appearing first in Eqs.  (\ref{c1n}) and (\ref{c2n}), are given by
\begin{eqnarray} \alpha&=& 2\,2^{\frac{1}{5}}\gamma^{12/5}\, \left[
3\,{{{\omega }_x}}^6 + 4\,{{{\omega }_y}}^4\,{{{\omega }_z}}^2 -
2\,{{{\omega }_y}}^2\,{{{\omega }_z}}^4\right.\\\nonumber &+&\left.
{{{\omega }_x}}^4\,( 3\,{{{\omega }_y}}^2 + 5\,{{{\omega }_z}}^2) +
2\,{{{\omega }_x}}^2\,( 2\,{{{\omega }_y}}^4 + {{{\omega }_z}}^4)
\right]\\\nonumber &/&\left({\pi }^{\frac{6}{5}}\,{{{\omega
}_x}}^4\,{{{\omega }_y}}^2\, \Omega_{xy}^4 \Omega_{xz}^4\right),
\end{eqnarray} \begin{eqnarray} \beta &=& -2\,2^{\frac{1}{5}}\gamma^{12/5},
\left[ 3\,{{{\omega }_x}}^8 - 2\,{{{\omega }_x}}^2\,{{{\omega
}_y}}^4\,{{{\omega }_z}}^2 \right.\\\nonumber &-&4\,{{{\omega
}_y}}^6\,{{{\omega }_z}}^2 + 2\,{{{\omega }_y}}^4\,{{{\omega }_z}}^4 +
{{{\omega }_x}}^6\,( 7\,{{{\omega }_y}}^2 + 2\,{{{\omega }_z}}^2
)\\\nonumber &+& \left.{{{\omega }_x}}^4\,( 6\,{{{\omega }_y}}^4 -
{{{\omega }_y}}^2\,{{{\omega }_z}}^2 + {{{\omega }_z}}^4 )
\right]\\\nonumber &/&\left({\pi }^{\frac{6}{5}}\, {{{\omega
}_x}}^4\,{{{\omega }_y}}^2\, \Omega_{xy}^4\, \Omega_{xz}^4\right),
\end{eqnarray} and \begin{equation} \eta=
\frac{2\,2^{\frac{1}{5}}\,\gamma^{12/5}\left(\Omega_{xy}^2\right) \, {
}^{\frac{12}{5}}}{{\pi }^{\frac{6}{5}}\,{{{\omega }_x}}^6\,{{{\omega
}_y}}^4\, \Omega_{xz}^4}, \end{equation} respectively.


\begin{references}

\bibitem{jin} D.  S.  Jin, J.  R.  Ensher, M.  R.  Mathews, C.  E.  Wieman,
and E.  A.  Cornell, Phys.  Rev.  Lett.  {\bf 77}, 420 (1996).

\bibitem{kett} M.-O.  Mewes, M.  R.  Andrews, N.  J.  van Druten, D.  S.
Durfee, C.  G.  Townsend, and W.  Ketterle, Phys.  Rev.  Lett.  {\bf 77},
988 (1996).

\bibitem{stringari_old} S.  Stringari, Phys.  Rev.  Lett.  {\bf 77}, 2360
(1996).

\bibitem{singh} K.  G.  Singh and D.  S.  Rokhsar, Phys.  Rev.  Lett.  {\bf
77}, 1667 (1996).

\bibitem{burnett} M.  Edwards, P.  A.  Ruprecht, K.  Burnett, R.  J.  Dodd,
and C.  W.  Clarck, Phys.  Rev.  Lett.  {\bf 77}, 1671 (1996).

\bibitem{castin} Y.  Castin and R.  Dum, Phys.  Rev.  Lett.  {\bf 77}, 5315
(1996).

\bibitem{perez} V.  M.  Perez-Carcia, H.  Michinel, J.  I.  Cirac, M.
Lewenstein, and P.  Zoller, Phys.  Rev.  Lett.  {\bf 77}, 5320 (1996).


\bibitem{stringari} D.  Gu\'ery-Odelin and S.  Stringari, Phys.  Rev.
Lett.  {\bf 83}, 4452 (1999).

\bibitem{pires} M.  O.  da C.  Pires and E.  J.  V.  de Passos, J.  Phys.
B:  At.  Mol.  Opt.  Phys.  {\bf 33}, 3929 (2000).

\bibitem{foot_first} O.  M.  Marag\`o, S.  A.  Hopkins, J.  Arlt, E.
Hodby, G.  Hechenblaikner, and C.  J.  Foot, Phys.  Rev.  Lett.  {\bf 84},
2056 (2000).


\bibitem{foot_beliaev} E.  Hodby, O.  M.  Marag\`o, G.  Hechenblaikner, and
 C.  J.  Foot, Phys.  Rev.  Lett.  {\bf 86}, 2196 (2001).

\bibitem{foot_temp} O.  M.  Marag\`o, G.  Hechenblaikner, E.  Hodby, and C.
J.  Foot, cond-mat/0101213.

\bibitem{eugen} B.  Jackson and E.  Zaremba, cond-mat/0105465.

\bibitem{foot_private} C.  J.  Foot, private communication.


\bibitem{pethick} G.  Baym and C.  J.  Pethick, Phys.  Rev.  Lett.  {\bf
76}, 6 (1996).


\bibitem{stringari_modern} F.  Dalfovo, S.  Giorgini, L.  P.  Pitaevskii,
and S.  Stringari, Rev.  Mod.  Phys.  {\bf 71}, 463 (1999).

\bibitem{beliaev} S.  T.  Beliaev, Sov.  Phys.  JETP {\bf 34}, 323 (1958).

\bibitem{foot_lower} See also G.  Hechenblaikner, E.  Hodby, S.  A.
Hopkins, O.  M.  Marag\`o, and C.  J.  Foot, cond-mat/0105394.



\end{references}
\end{document}